# 'Team-in-the-loop': Ostrom's IAD framework 'rules in use' to map and measure contextual impacts of AI

Deborah Morgan[1,2], Youmna Hashem[1], John Francis[1], Saba Esnaashari[1], Vincent J Straub[1], Jonathan Bright[1]

1 Public Policy Programme, Alan Turing Institute, London, NW1 2DB, UK
2 ART-AI, Department of Computer Science, University of Bath, Bath, BA2 7AY, UK

**Abstract:** This article explores how the 'rules in use' from Ostrom's Institutional Analysis and Development Framework (IAD) can be developed as a context analysis approach for AI. AI risk assessment frameworks increasingly highlight the need to understand existing contexts. However, these approaches do not frequently connect with established institutional analysis scholarship. We outline a novel direction illustrated through a high-level example to understand how clinical oversight is potentially impacted by AI. Much current thinking regarding oversight for AI revolves around the idea of decision makers being 'in-the-loop' and, thus, having capacity to intervene to prevent harm. However, our analysis finds that oversight is complex, frequently made by teams of professionals and relies upon explanation to elicit information. Professional bodies and liability also function as institutions of polycentric oversight. These are all impacted by the challenge of oversight of AI systems. The approach outlined has potential utility as a policy tool of context analysis aligned with the 'Govern and Map' functions of the National Institute of Standards and Technology (NIST) AI Risk Management Framework; however, further empirical research is needed. Our analysis illustrates the benefit of existing institutional analysis approaches in foregrounding team structures within oversight and, thus, in conceptions of 'human in the loop'.

# 1   Introduction

We propose the use of the 'rules in use' from Ostrom's Institutional Analysis and Development (IAD) (Ostrom, 2005, 2011) as an analytical tool to understand existing contexts of AI deployment in the public sector. Such work supports developing approaches to AI risk management such as the National Institute of Standards and Technology (NIST) AI Risk Management Framework (NIST AI RMF, 2023) to understand existing context and the potential impacts of artificial intelligence (AI) systems in complex public sector settings. AI systems in the public sector have significant potential to impact service delivery through predictive modelling, early detection of risk, and the wider use of 'real-time transactions data' (Margetts, 2022). However, the distinct context and inherent complexity of public sector deployment necessitates reliable operational protocols, transparent and explainable processes, alongside alignment with the normative expectations of the setting (Straub et al., 2023). Understanding the expectations and rules that structure these settings is particularly vital for AI deployment in 'high stakes' public sector decision-making such as medical diagnosis, service allocation, or educational assessment (Rudin, 2019; Busuioc, 2021; Smith, 2021) where mistakes can result in individual and collective harms (Sambasivan *et al.,* 2021; Smuha*,* 2021).

Developing AI standards, guidance, and risk frameworks (ISO/IEC 42001: 2023, NIST AI RMF) are increasingly referenced as best practices to assure AI use against; for example, the United States NIST AI Risk Management Framework and accompanying playbook (NIST AI RMF) is noted within recent AI Assurance guidance issued by the UK Government (DSIT, 2024). The NIST framework aims to support organisations through four distinct, yet interconnected, functions which 'govern, map, measure, and monitor AI development and deployment' across the AI lifecycle. Within the crosscutting Govern function, the framework highlights the need to 'connect AI governance to existing organizational governance and risk controls' as a suggested action for the governance of AI systems (Govern 1.2, NIST AI RMF Playbook). The Map function emphasises the need for context mapping and analysis to establish and understand the intended context of use (Map 1.1, NIST AI RMF Playbook). This includes the need to 'plan for risks related to human-AI configurations, and document requirements, roles, and responsibilities for human oversight of deployed systems'.

Various frameworks and approaches are proposed for such risk and context mapping (SIMLab,2017) which are often aimed at the mapping of risks and impacts of the whole AI system in a broad context rather than targeted at an operational use case or at understanding specific governance dimensions, for example, oversight or the use of bureaucratic discretion. In addition, proposals are frequently situated at organisational governance decisions around deployment rather than as a mechanism to understand the complexity and polycentric nature of an operational context. Recent economic analysis of AI in Radiology noted that 'the use of AI in practice will be mediated by organizational incentives and the regulatory environment' (Agarwal *et al.,* 2023). Institutional analysis applying existing tools, such as the IAD, can make a significant contribution to understanding the complexity of such organisational incentives, existing contexts, and applicable regulatory processes to situate and anticipate AI impacts.

Human oversight is defined as 'the inclusion of a means to ensure shared awareness of the situation, such that the person making the decisions has sufficient information at the time they must intervene', applying an adapted version of Dignum's definition of human control (Dignum, 2019). Establishing the boundaries of such 'shared awareness' and of 'sufficient information' are significant challenges within oversight of decisions. Oversight is the focus of the example used here as the boundaries, form, and definitions of 'human in the loop' oversight of AI are a core concern of emerging governance proposals (see Article 14 of the European Commission proposed Regulation on Artificial Intelligence, 2021). However, the forms and triggers of such human oversight are not yet settled in many contexts of deployment (Sterz *et al*, 2024). Oversight of high-stakes public sector decisions is frequently contextual and conducted by teams of professionals acting within specific institutional 'norms' or defined rules of practice (Morgan *et al.,* 2022). Examples of such processes include the additional questions asked and noted by an experienced consultant or a multi-disciplinary team meeting to collectively determine treatment plans. Such team processes represent a 'normative expectation' (Straub *et al.,* 2023) developed to support safe deployment and engender public trust. Detailed understanding of this institutional context is increasingly necessary for policy analysts developing AI assurance proposals and safety mechanisms alongside AI development teams conceptualising a system.

We begin with a brief review of literature regarding AI, decision making and oversight. We then apply the IAD rules to a high-level clinical example in which we conceptualise oversight as an 'interaction setting' (Klok and Denters, 2018) and then problematise potential impacts of AI on this. This overview is inevitably broad and seeks to present a potential research direction rather than a definitive model. We conclude with a discussion of the implications and utility of this approach and suggest future research directions to connect developing AI assurance with existing forms of institutional analysis.

## 2. Decision-making and AI

High-stakes decisions within the public sector are frequently made and overseen at an operational level within an institutional, often hierarchical, team aligned to formal and informal rules of behaviour. Within such systems, delegated decision-making is an established norm of practice providing benefits in utilisation of time (Harris and Raviv, 2002), especially in dynamic environments (Radner, 1993). The reasons for existing hierarchical structures of delegation, or 'vertical specialisation', include the need for team coordination, and the integration of professional expertise to permit 'the operative personnel to be held accountable for their decisions' (Simon, 1997). Diverse fields have explored such hierarchical-based processes and the limitations of them to produce optimal societal (Milanovich *et al.,* 1998; Sydor *et al.,* 2013; Bould *et al.,* 2015) or business outcomes (Christensen, 2016). Within a healthcare context, specific concerns have been raised regarding the flow of information, for example the impact of 'authority gradient' upon junior doctors (Cosby, 2004; Peadon, Hurley and Hutchinson, 2020). However, such processes remain a dominant paradigm of the clinical decision-making context.

Emerging risk management frameworks and toolkits emphasise the need to engage with domain experts and end users when developing an AI system to understand the existing context of a proposed application (see Map 1.1 NIST AI RMF and ISO/IEC 42001, Clause 4). The emphasis upon shared awareness in the definition of oversight, detailed in Section 1, acknowledges that relevant, yet potentially unavailable, information from different stages of AI development (e.g., the origin and extent of the dataset, Paullada *et al.,* 2021) are all of relevance in establishing the level of 'sufficiency' required from decision-makers. Questions of transparency, knowledge and cross-disciplinary understanding may all potentially inhibit the capability of the designated 'human in the loop' to establish a level of sufficiency of information necessary to intervene or to query a system in use.

Processes and theories of human oversight or 'meaningful' control in relation to automation originated within safety critical systems approaches and human factors work (Cacciabue, 2004; Savioja and Norros, 2013). In such contexts, the human provides a fail-safe mechanism in the event of a failure. Research and proposed mechanisms of 'human in the loop' oversight of wider automated systems have been studied within specific domains including defence (Tyworth, *et al*., 2013; Santoni de Sio and van den Hoven, 2018) and data protection (Binns and Veale, 2021). More recent advances in digital and AI technologies have generated burgeoning literature examining the challenges and effectiveness of actual human oversight (Green, 2022; Sterz *et al.,* 2024) of complex, opaque AI systems.  Within the clinical context, there are an increasing range of AI systems to support decision making particularly within the field of medical imaging classification (Rashidi *et al.,* 2021), diagnosis support (Cai *et al.,* 2019) and risk prediction models to identify high-risk patients (Clift *et al.,* 2020; Giordano *et al.,* 2021). AI assurance guidance is emerging alongside these advances with proposed AI clinical trial models and audits (Rivera *et al.,* 2020; Liu *et al.,* 2022). However, unlike established clinical assurance processes such as randomised clinical trials or ex post treatment monitoring protocols these approaches are not yet integrated or mandated.

**3. Institutional Analysis and Development Framework and the 'rules in use'**

Elinor Ostrom developed the IAD Framework (IAD) to analyse and understand institutions, defined broadly as 'the prescriptions humans use to organise all forms of repetitive and structured interactions' (Ostrom, 2005). The focal point of the IAD is an action arena, comprising actors within action situations affected by external variables (Ostrom, 2005; Clement, 2010). In the action arena actors engage through patterns of interaction to create outcomes. Undertaking a full IAD analysis for a distinct AI context and application would be an interdisciplinary task requiring input and participation from diverse stakeholders to understand the 'rules in use' and the interdependencies of various institutions. In this article, we aim to sketch the outlines of such work through a high-level example applying one dimension of the IAD. Further work to refine, model and assess the limitations of the use of the analysis outlined is necessary alongside developing a range of contextual examples.

The 'rules in use' of a situation condition these relevant interactions (Cole, 2014). The IAD classifies rules into seven categories, (Kiser and Ostrom, 1982; Ostrom, 1986, 2005) as detailed in Figure 1. They are the "dos and don'ts' that one learns on the ground that may not exist in any written document' (Ostrom, 2005). Ostrom highlighted the applicability of institutional analysis to different settings, including team structures (Ostrom, 1985) and it has subsequently been applied to team analysis of software developers (Tenenberg, 2008) and digital workers (Piciocchi *et al.,* 2019).

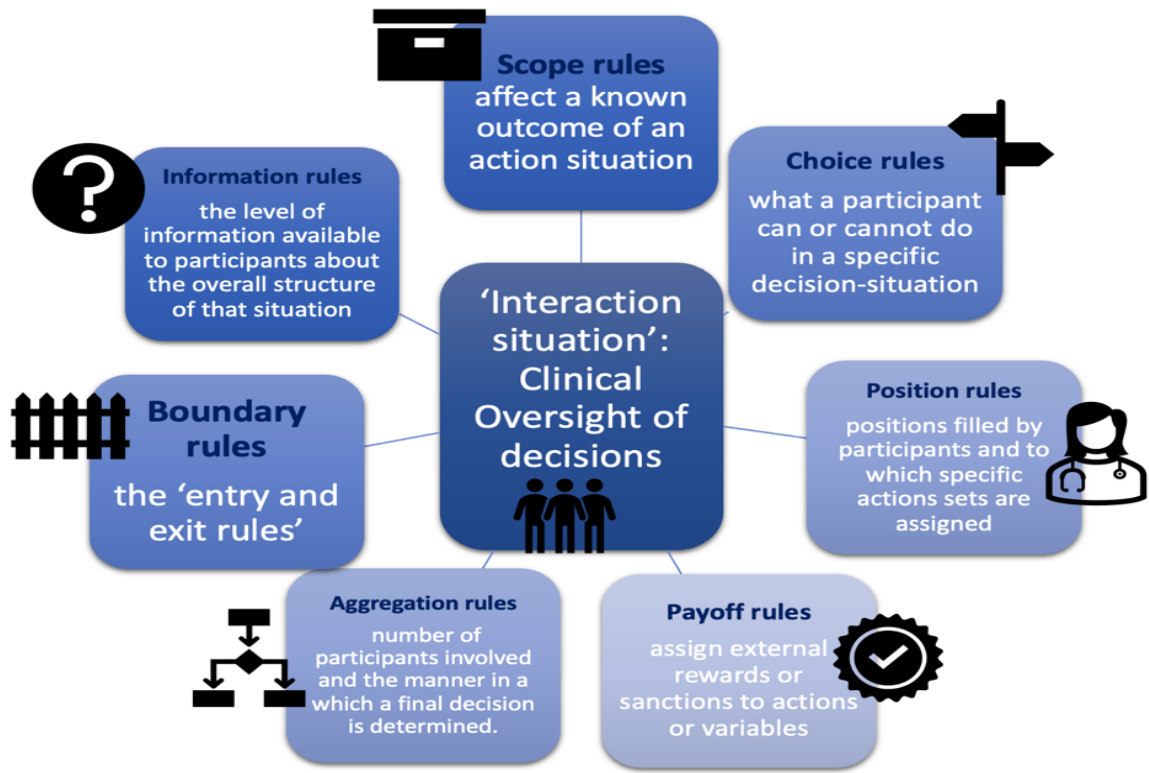

Figure 1: Definitions of the 'rules in use' adapted from Ostrom's IAD framework (Ostrom, 2005)

## 4. The 'rules in use' of clinical oversight

The example used is situated within England and Wales and the institutions aligned to healthcare provision are referred to throughout as the National Healthcare Service (NHS). Healthcare is selected for analysis due to its complexity, the range of clinical AI systems in development and literature regarding applications and impacts; however, the approach outlined has utility to a variety of public sector decision contexts. In the review of each rule below, we outline significant dimensions of the existing context and then suggest potential impacts of AI. The aim is not to provide a comprehensive example or model but to highlight the need for further theoretical and empirical analysis of the utility of existing policy frameworks, such as the IAD, to support detailed analysis of AI impact and existing context. We include relevant aspects of the NIST AI RMF to align this approach with an emerging AI assurance method. The data for this analysis is drawn primarily from academic and grey literature with some references to the limited available sources of empirical data on AI use and user experiences.

### 4.1. Position rules: - *positions filled by participants and to which specific actions sets are assigned*

Positions within a clinical team are often nested within a hierarchy spanning doctors, or fellow clinical professionals, in training through to senior consultants who may also hold clinical management or research positions (BMA, 2022). The actions allotted to a position are closely aligned to professional training requirements. The role and scope of senior members of the team to impact progression through assessment is a significant element within the development of oversight skills and process. A direct example can be seen within the postgraduate foundation year of training when junior doctors are required to complete a series of recorded Supervised Learning Events (SLEs) (UK Foundation Programme, 2021) as a record of feedback from a supervisor of direct observation of procedures. Liability is also aligned to a professional position and attributed to the person exercising 'skilled care' appropriate to a reasonable body of opinion of other practitioners in that field, (defined as the *Bolam* test following Bolam v Friern Hospital Management Committee [1957]*).

The role of actors features significantly within the NIST AI RMF Map 1.1 function which details the need for AI actors to work collaboratively and to use context mapping to understand risks relating to end users and context-specific impacts such as legal requirements. The position rules support such mapping to delineate the authority and decision-making powers of existing operational actors. In clinical decision-making, professionals operate within teams of adjacent professionals, with shared norms of clinical oversight and review. Placing data scientists and informaticians in senior positions could support necessary questioning and scrutiny of AI systems alongside clinicians in the multi-disciplinary team meeting (NHS AI Lab & Health Education England, 2022). However, the boundaries and positions of such digital clinical professionals is unclear at present. The NHS AI Lab & Health Education England report (2022), which explored the factors influencing healthcare workers' confidence in AI, highlighted that retention of digital professionals was a challenge in part due to the remuneration levels classifying their pay scale at 'administrative/clerical' staff level. They also noted that there are few individuals in senior positions, and these are not represented within high-level decision-making (NHS AI Lab & Health Education England*,* 2022).

### 4.2. Boundary rules: *the 'entry and exit rules'*

Boundary rules necessarily vary across different disciplines; however, many have defined training pathways and requirements established by professional regulators such as the General Medical Council (GMC) and domain specific Royal College requirements for postgraduate progression. Doctors are licensed, follow defined practice standards of professional behaviour and are subject to disciplinary assessment in the event of a complaint. Clinicians also follow boundary rules which underpin their professional competencies, including, the maintenance of training records, and completing a minimum number of hours of work or attending/providing training sessions (see GMC *Excellence by design,* 2017). Therefore, a range of polycentric decision-making bodies including professional bodies, medical trade unions, Royal

Colleges, institutional, regional, and national training bodies, and tribunal disciplinary bodies all influence and structure the rules of entry and exit of clinical professionals.

Medical legal precedent has established the boundaries of liability and pharmaceuticals, and medical devices are regulated, externally verified, and monitored by external regulatory agencies and processes. Various proposals exist and are in development for clinical audits and potential clinical trials of AI systems (Rivera *et al.,* 2020; Liu *et al.,* 2022). However, their use and integration in routine clinical practice is not yet clear nor is the place of 'hybrid' medical and data professionals within clinical teams. Progression through medical training also provides structured opportunities to gain skills and oversight capabilities. The removal of aspects of training experience through automation or augmentation of parts of the decision making 'stack' may then impact the subsequent capability of 'downstream professionals' (Schwartz *et al.,* 2022) to oversee a system if that task is abstracted away (Choudhury and Chaudhry, 2024). Automation or reduction of human inputs may not be significant for the performance of a decision itself, if clinical performance is reliably verified, but may impact operational oversight to detect outliers, model drift or wider systemic issues. The boundary rules support mapping and contextual understanding ahead of deployment in alignment with the overall function Map 1 of the NIST AI RMF to establish and understand context. They also provide understanding of a significant aspect of the governance context in alignment with Govern 1.1 of the NIST AI RMF as professional standards, liability and defined training are a core component of operational governance in a clinical decision-making context.

**4.3. Aggregation rules:** *number of participants involved and the manner in a which a final decision is determined.*

Different medical specialities interact with clinical colleagues within multi-disciplinary teams and work alongside patients to reach shared decisions regarding treatment (NHS England and NHS Improvement, 2019). The clinical hierarchical situation is a 'nonsymmetric aggregation rule' (Ostrom, 2005) in that participants are treated differently according to seniority and position held. Each admitted patient is allocated a named 'responsible' consultant which is an example of aggregation of responsibility in a senior individual within a wider clinical team. The increasing utilisation of multi-disciplinary teams and vital role of patient shared decision making are all additional inputs and structure the final aggregated clinical decision (Makoul and Clayman, 2006; Elwyn *et al.,* 2010; NHS England and NHS Improvement, 2019). The liability of multidisciplinary teams has been explored by Klemm and Lehman, (2021) who highlight that each practitioner potentially retains liability for their own disciplinary area within a collective decision. The accountability for a clinical decision may lie with a senior party; however, a junior party or fellow clinical professionals will also partially hold liability appropriate to their role and status through professional obligations. Best practice 'norms', good management practice and position rules regarding ethical and legal actions all impact the range of actions and way in which aggregation is conducted in practice.

The form in which a clinical decision is aggregated is context and domain specific. Each decision will combine varying levels of individual, team, patient, and institutional elements. The ability to check, review and 'disaggregate' the processes underpinning a decision through information exchange with other clinical practitioners is integral. Whilst AI systems clearly do not prevent or change such processes, they may contribute to a premature closing of options through algorithmic deference or automation bias (Alon-Barkat and Busuioc, 2022; Strauß, 2021; Academy of Medical Royal Colleges, 2019). Work exploring human interaction with decision-recommendation systems has found some evidence of algorithmic deference and of struggles to evaluate systems and risk assessments (Green, 2022; Green and Chen, 2021). The NIST AI RMF function Map 2.2 requires information about how the AI systems knowledge limits and process of human oversight are documented. This function highlights the risks explored above in that, 'downstream decisions can be influenced by end user over-trust or under-trust, and other complexities related to AI-supported decision-making'. Suggested actions include testing human-AI collaborations and documentation regarding dependencies of systems on upstream data and other AI systems. The aggregation rules highlight the relational nature of team decision making processes which is a significant factor in considerations of documentation and methods to enable confirmatory review and disaggregation of decision-making to reassure a position holder.

**4.4. Information rules:** *the level of information available to participants about the overall structure of that situation*

Dialogue and verbal communication, alongside recording of reasoning, are vital to elicit relevant information within a high-stakes environment, particularly within critical situations (see work exploring the 'medical handover' (Bhabra *et al.,* 2007; Swain *et al.,* 2021). Within hierarchical team structures, achieving optimal information flow can be a challenging area as authority differentiation may impact the ability of junior team members to share opinions, and even facts, relevant for necessary oversight of delegated actions (Brennan and Davidson, 2019; McMaster, Phillips and Broughton, 2016; Francis, 2015). Senior members will often question junior team members thoroughly to test knowledge and decision recommendations provided. The manner and style of questioning can be critiqued as it may contribute to low inclusivity and reluctance for junior team members to provide reciprocal oversight of senior decisions (McMaster, Phillips and Broughton, 2016). However, the ability to provide robust information, indicate uncertainty and answer questions outlining the variables and reasoning underpinning a decision is a component of clinical oversight. Additional aspects of information rules that impact oversight are the frequency and accuracy of communication as there may need to be regular 'check ins' with limited time available in which to discuss a case. Decisions may also be inherited but then enacted by a junior doctor or fellow clinician (Bull, Mattick and Postlethwaite, 2013; Dusse *et al.,* 2021). Information rules are also particularly relevant as audit procedures to log the reasons underpinning treatment decisions flows.

Information flows are dynamic, situational, and adaptive between members of a clinical team. The ability to seek further advice from senior colleagues regarding a challenging or uncertain situation is integral within medicine, and more broadly within professional contexts. The ability of medical professionals to balance their clinical judgement against AI inputs is a potential operational challenge. The risks of 'infobesity' (Methnani *et al.,* 2021) whereby a human's cognitive ability is overloaded and algorithmic deference whereby a human defers to an output (Green and Chen, 2019) both require live assessment, testing and external validation metrics in simulated operational contexts to assess actual prevalence and risk. The mechanisms through which information are captured and recorded are detailed extensively across the NIST AI RMF. For example, NIST AI RMF Function Map 2.2 recommends that 'documentation provides sufficient information to assist relevant AI actors when making informed decisions and taking subsequent actions'. However, as noted above the integration of such information and documentation for operational actors requires awareness of the risks of 'infobesity' to pay careful attention to how information is actually integrated and used by position holders to review AI outputs.

**4.5. Payoff rules:** *assign external rewards or sanctions to actions or variables.*

Failure to oversee a decision correctly can incur sanctions to a clinician, their institutional reputation and potentially investigations by professional bodies. Conversely oversight of junior clinicians can enhance the knowledge base of senior staff, support their advancement and is one dimension of delegated decision making (Simon, 1997). The GMC's 'Good Medical Practice' commitments make care duties explicit through statements such as: 'Make the care of your patients your first concern' and 'Act promptly if you think that patient safety or dignity may be seriously compromised' (GMC, 2024)  At an individual level, a junior member of the team may be directly affected by how they conduct oversight requests as their training and progression are partially dependent upon senior team members' perceptions and assessments of them (Bull, Mattick and Postlethwaite, 2013).

Within a team, delegation supports workload management. For junior team members, this context, ideally, provides training, oversight, and support as they progress. The incentives to incorporate, 'train' and oversee novel AI systems within a team may not be as prevalent for individual clinicians when compared to the pre-existing institutional and normative professional incentives to train and support junior colleagues. Mistakes and errors may also be less tolerated and be harder to identify and understand in situ from AI outputs. Questioning and shared contexts provide easily understood clues and shared understanding regarding the reasons for mistakes. They could be then easier to address promptly in situ than an error made through incorrect specification, flawed data, or model drift from AI systems. Various aspects of the NIST AI RMF suggest policies and procedures to report risks and incidents from AI systems including function Govern 4.3 which recommends established processes to report incidents and Govern 6.2 which suggests establishing a process for third parties to report potential vulnerabilities, risks, or biases. However, the payoff rules illustrate the significance of the normative social and professional team context in the delegation and oversight of decisions.

**4.6. Scope rules:** *affect a known outcome of an action situation*

Scope rules are those which affect a known, potential outcome of an action situation. The scope rules 'delimit the range of possible outcomes, In the absence of a scope rule, actors can affect any physically possible outcomes' (Cole, 2014). Scope rules in a clinical oversight situation impact the range of treatment options decided upon by teams of clinicians and patients. The range of possible outcomes in a clinical context will necessarily be impacted by institutional and national guidance on treatment pathways, for example the guidance contained within the UK National Institute for Health and Care Excellence (NICE) Clinical Knowledge Summaries. Practice guidance, medical literature and training curricula comprise the scope rules of potential treatment options within the boundaries of oversight and responsibility structured by professional standards, training responsibilities and medical liability precedents. However, wider institutional or service level performance requirements and budget constraints may also impact the range and form of scope rules in practice.

As scope rules structure an outcome it may appear that AI systems used as tools to achieve a particular defined treatment outcome may not impact scope rules significantly. However, as discussed, the impact of existing, trusted guidance, colleague discussions and evaluation processes on structuring clinical decisions is significant and careful consideration is needed to provide trusted guidance and approaches regarding the future integration and use of AI recommendations. At an operational level, a conflict between an AI recommendation of a part of the scope of a treatment pathway and a doctor's decision may raise significant questions of clinical practice, autonomy and of liability in the event of later concerns. Again, the need for trusted and verified process of domain evaluation and valuation to support doctors in interacting and verifying AI inputs is significant. As noted within the work undertaken by the NHS AI Lab & Health Education England (2022), guidance regarding individual and organisational liability is necessary to support operational professionals in developing appropriate confidence regarding the use of AI systems.

**4.7 Choice rules:** *what a participant can or cannot do in a specific decision-situation*

Choice rules outline what a participant can or cannot do in a specific decision-situation and are also described as authority rules. Such rules could impact oversight if a more senior party was not available to oversee decisions or may be affected by the extent to which a junior colleague is 'trusted' by a senior party (Kennedy *et al.,* 2007). Choice rules structure the action capacity of an individual. As noted in the rules relating to position and boundary the actions to take high-risk clinical decisions are underpinned by a process of professional training overseen by a professional regulator within boundaries of professional responsibility. Legal precedent has established clinical liability aligned to a reasonable standard of care. Whilst clearly the choice to take an action is aligned to position, the operation of this rule is contextual depending upon an actor's position and capacity within the clinical team.

AI systems within decision making may impact action choice in a range of ways by expanding the range of potential treatment options, providing a wider range of data for consideration or they may suggest treatment options based on a multiplicity of data sources. As noted, the ability and capacity to oversee a decision is a significant element of the operation of clinical oversight. Clear definitions of the role and boundaries of an AI system in the decision-making processes is a significant component of the NIST AI RMF spanning governance, mapping of context and management of the system in use. At present professional standards, guidance and structures delineate the choices available to a decision maker often nested within a team structure. The impact of AI systems within decision making upon these choices may be challenging to measure and test individually; however, further work to explore team assessment or analysis of AI inputs which is more aligned to the team structures within clinical setting may be a potential area of further AI assessment and assurance research.

## 6      Discussion and Conclusion

This high-level example highlights the role of professional clinical training requirements and responsibilities alongside the importance of explanation and questioning within teams as confirmatory checks to reassure the oversight party. These provide multiple data sources and tests to generate the level of 'sufficient information' required. Delegation of decisions also facilies access to necessary experience and develops the profession within supervised defined boundaries. Professional bodies and regulators operate as institutional mechanisms of oversight through standards, training requirements and professional accountability. All of these components of oversight are potentially impacted and reconfigured by AI systems.  Work to understand AI use and to map context is increasingly necessary with the developing availability of AI systems notably generative AI models. Recent survey research has found that 45% of UK public sector professional respondents were aware of generative AI usage within their area of work, with 22% actively using a generative AI system; however, only a minority (32%) felt that there was clear guidance around the use of generative AI in their workplace (Bright *et al.,* 2024). We consider that the IAD framework has much to offer in system analysis and mapping of determinative polycentric institutions particularly in its focus upon the roles and actions of relevant position holders within a defined 'action' context. We have situated this analysis within a high-level clinical example; however, it does suggest areas of challenge more broadly in balancing expert knowledge with AI inputs to provide oversight within professional teams. Such work is an increasing priority to investigate how 'human-AI teams should be configured to reduce the likelihood of negative impacts/ harms to individuals, groups, communities, and society' (NIST, 2023).

Our analysis also suggests the more appropriate description of 'team in the loop' to reflect the operational reality of existing high stakes oversight and decision-making structures. It is also reflective of the need for diverse 'teams' with contextual and professional expertise to conduct algorithmic audits and to oversee AI systems in use within high-stakes domains as noted within the NIST AI RMF (Kazim, Denny and Koshiyama, 2021; Liu *et al.,* 2022; Raji *et al.,* 2022). Adjacent public sector contexts such as education, defence, social care provision and public administration potentially all share the risks of deskilling, 'infobesity' and rely upon dialogue and questioning within decision oversight. Whilst each sector has a different configuration of the 'rules in use' within an action arena, there are commonalities in balancing expertise with decision support from AI systems to develop appropriate confidence. Recognising and mapping these existing structures of oversight within high-stake contexts highlights the primacy of teams of professionals operating within existing rules that structure and bound human oversight. Oversight of AI systems within such processes is more accurately and necessarily a domain specific 'team in the loop' endeavour. Recognising and understanding the form and rules of such collective oversight and the situation complexity is a necessary step for policy and regulatory workstreams to develop the professionals, standards, systems, and training needed to engage with operational integration of AI systems. Professional standards, accountability and oversight have been developed and evolved overtime to support high standards of care. Public trust and confidence in high stakes public sector decision making will not be secured 'simply by putting a human in the loop', but rather these issues will require entire teams or systems to be considered in the loop of oversight. (Binns, 2022)

**Acknowledgements:** This work was supported by Towards Turing 2.0 under the EPSRC Grant EP/W037211/1 and The Alan Turing Institute. Deborah Morgan is a PhD researcher in the Accountable, Responsible and Transparent AI CDT in the Department of Computer Science at the University of Bath supported by UKRI Grant EP/S023437/1.

# References

Academy of Medical Royal Colleges (2019) *Artificial Intelligence in Healthcare*. Available at: http://www.aomrc.org.uk/reports-guidance/artificial-intelligence-in-healthcare/.

Agarwal, N. *et al*. (2023) 'Combining Human Expertise with Artificial Intelligence: Experimental Evidence from Radiology'. Rochester, NY. Available at: https://papers.ssrn.com/abstract=4505053 (Accessed: 6 June 2024).

Alon-Barkat, S. and Busuioc, M. (2022) 'Human–AI Interactions in Public Sector Decision Making: "Automation Bias" and "Selective Adherence" to Algorithmic Advice', *Journal of Public Administration Research and Theory*, p. muac007. Available at: https://doi.org/10.1093/jopart/muac007.

Arnold, C. (2023). Inside the nascent industry of AI-designed drugs. *Nat Med* **29**, 1292–1295. Available at: https://doi.org/10.1038/s41591-023-02361-0

Bhabra, G. *et al*. (2007) 'An Experimental Comparison of Handover Methods', *The Annals of The Royal College of Surgeons of England*, 89(3), pp. 298–300. Available at: https://doi.org/10.1308/003588407X168352.

Binns, R. and Veale, M. (2021) 'Is that your final decision? Multi-stage profiling, selective effects, and Article 22 of the GDPR', *International Data Privacy Law*, 11(4), pp. 319–332. Available at: https://doi.org/10.1093/idpl/ipab020.

*Bolam v Friern Hospital Management Committee [1957] 1 W.L.R. 582 (26 February 1957) Practical Law*. Available at: http://uk.practicallaw.thomsonreuters.com/D-016-0979?transitionType=Default&contextData=(sc.Default)&firstPage=true

Bould, M.D. *et al*. (2015) 'Residents' reluctance to challenge negative hierarchy in the operating room: a qualitative study', *Canadian Journal of Anesthesia/Journal canadien d'anesthésie*, 62(6), pp. 576–586. Available at: https://doi.org/10.1007/s12630-015-0364-5.

Brennan, P.A. and Davidson, M. (2019) 'Improving patient safety: we need to reduce hierarchy and empower junior doctors to speak up', *BMJ*, p. l4461. Available at: https://doi.org/10.1136/bmj.l4461.

Bright, J. *et al*. (2024) 'Generative AI is already widespread in the public sector'. arXiv. Available at: http://arxiv.org/abs/2401.01291

British Medical Association (2022) 'Toolkit for doctors new to the UK Doctors' titles explained' Available at: https://www.bma.org.uk/advice-and-support/international-doctors/life-and-work-in-the-uk/toolkit-for-doctors-new-to-the-uk/doctors-titles-explained

Bull, S., Mattick, K. and Postlethwaite, K. (2013) '"Junior doctor decision making: isn't that an oxymoron?" A qualitative analysis of junior doctors' ward-based decision-making', *Journal of Vocational Education & Training*, 65(3), pp. 402–421. Available at: https://doi.org/10.1080/13636820.2013.834955.

Busuioc, M. (2021) 'Accountable Artificial Intelligence: Holding Algorithms to Account', *Public Administration Review*, 81(5), pp. 825–836. Available at: https://doi.org/10.1111/puar.13293.

Cacciabue, P.C. (2004) *Guide to applying human factors methods*. London: Springer.

Cai, C.J. *et al.* (2019) 'Human-Centered Tools for Coping with Imperfect Algorithms During Medical Decision-Making', in *Proceedings of the 2019 CHI Conference on Human Factors in Computing Systems. CHI '19: CHI Conference on Human Factors in Computing Systems*, Glasgow Scotland Uk: ACM, pp. 1–14. Available at: https://doi.org/10.1145/3290605.3300234.

Choudhury, A. and Chaudhry, Z., 2024. Large Language Models and User Trust: Consequence of Self-Referential Learning Loop and the Deskilling of Health Care Professionals. *Journal of Medical Internet Research*, *26*, p.e56764.

Christensen, C.M. (2016) *The innovator's dilemma: when new technologies cause great firms to fail*. Paperback. Boston, Massachusetts: Harvard Business Review Press (The management of innovation and change series).

Clement, F. (2010) 'Analysing decentralised natural resource governance: proposition for a "politicised" institutional analysis and development framework', *Policy Sciences*, 43(2), pp. 129–156. Available at: https://doi.org/10.1007/s11077-009-9100-8.

Clift, A.K. *et al.* (2020) 'Living risk prediction algorithm (QCOVID) for risk of hospital admission and mortality from coronavirus 19 in adults: national derivation and validation cohort study', *BMJ*, p. m3731. Available at: https://doi.org/10.1136/bmj.m3731.

Cole, D.H. (2014) 'Formal Institutions and the IAD Framework: Bringing the Law Back In', *SSRN Electronic Journal* [Preprint]. Available at: https://doi.org/10.2139/ssrn.2471040.

Cosby, K.S. (2004) 'Profiles in Patient Safety: Authority Gradients in Medical Error', *Academic Emergency Medicine*, 11(12), pp. 1341–1345. Available at: https://doi.org/10.1197/j.aem.2004.07.005.

Department for Science, Innovation & Technology. (2024). *Guidance. Introduction to AI assurance*. Available at: https://www.gov.uk/government/publications/introduction-to-ai-assurance/introduction-to-ai-assurance

*Developing healthcare workers confidence in AI | Health Education England* (no date) *Health Education England | Digital Transformation*. Available at: https://digital-transformation.hee.nhs.uk/building-a-digital-workforce/dart-ed/horizon-scanning/developing-healthcare-workers-confidence-in-ai (Accessed: 13 March 2023).

Dignum, V. (2019) *Responsible Artificial Intelligence: How to Develop and Use AI in a Responsible Way*. 1st edn. Springer Cham (Artificial Intelligence: Foundations, Theory, and Algorithms). Available at: https://doi.org/10.1007/978-3-030-30371-6.

Dusse, F. *et al.* (2021) 'Completeness of the operating room to intensive care unit handover: a matter of time?', *BMC Anesthesiology*, 21(1), p. 38. Available at: https://doi.org/10.1186/s12871-021-01247-3.

Elwyn, G. *et al.* (2010) 'Implementing shared decision making in the NHS', *BMJ*, 341(oct14 2), pp. c5146–c5146. Available at: https://doi.org/10.1136/bmj.c5146.

Proposal for a REGULATION OF THE EUROPEAN PARLIAMENT AND OF THE COUNCIL LAYING DOWN HARMONISED RULES ON ARTIFICIAL INTELLIGENCE (ARTIFICIAL INTELLIGENCE ACT) AND AMENDING CERTAIN UNION LEGISLATIVE ACTS. COM/2021/206 final. (2021). Available at: https://eur-lex.europa.eu/legal-content/EN/TXT/?uri=celex%3A52021PC0206

Francis, R. (2015) *Freedom to speak up: An independent review into creating an open and honest reporting culture in the NHS*. Freedom to speak up. Available at: https://webarchive.nationalarchives.gov.uk/ukgwa/20150218150512/http://freedomtospeakup.org.uk/the-report/

General Medical Council, (2017) Excellence by design. Available at: https://www.gmc-uk.org/education/standards-guidance-and-curricula/standards-and-outcomes/excellence-by-design

General Medical Council, (2024) Good medical practice. Available at: https://www.gmc-uk.org/professional-standards/professional-standards-for-doctors/good-medical-practice

Giordano, C. *et al.* (2021) 'Accessing Artificial Intelligence for Clinical Decision-Making', *Frontiers in Digital Health*, 3, p. 645232. Available at: https://doi.org/10.3389/fdgth.2021.645232.

Green, B. (2022) 'The flaws of policies requiring human oversight of government algorithms', *Computer Law & Security Review*, 45, p. 105681. Available at: https://doi.org/10.1016/j.clsr.2022.105681.

Green, B. and Chen, Y. (2019) 'The Principles and Limits of Algorithm-in-the-Loop Decision Making', *Proceedings of the ACM on Human-Computer Interaction*, 3(CSCW), pp. 1–24. Available at: https://doi.org/10.1145/3359152.

Green, B. and Chen, Y. (2021) 'Algorithmic Risk Assessments Can Alter Human Decision-Making Processes in High-Stakes Government Contexts', *Proceedings of the ACM on Human-Computer Interaction*, 5(CSCW2), pp. 1–33. Available at: https://doi.org/10.1145/3479562.

Harris, M. and Raviv, A. (2002) 'Organization Design', *Management Science*, 48(7), pp. 852–865. Available at: https://doi.org/10.1287/mnsc.48.7.852.2821.

Health Education England (no date) 'Enhancing supervision for postgraduate doctors in training'. Available at: https://www.hee.nhs.uk/enhancing-supervision.

International Organization for Standardization [ISO] (2023) *ISO/IEC 42001: 2023 Information technology — Artificial intelligence — Management system*, ISO [Online]. Available at https://www.iso.org/standard/74046.html.

Kazim, E., Denny, D.M.T. and Koshiyama, A. (2021) 'AI auditing and impact assessment: according to the UK information commissioner's office', *AI and Ethics*, 1(3), pp. 301–310. Available at: https://doi.org/10.1007/s43681-021-00039-2.

Kennedy, T.J.T. *et al.* (2007) 'Clinical Oversight: Conceptualizing the Relationship Between Supervision and Safety', *Journal of General Internal Medicine*, 22(8), pp. 1080–1085. Available at: https://doi.org/10.1007/s11606-007-0179-3.

Kiser, L. and Ostrom, E. (1982) 'The three worlds of action: A metatheoretical synthesis of institutional approaches in strategies of political inquiry', *Strategies of Political Inquiry; Ostrom, E., Ed.; Sage: Beverly Hills, CA, USA*, pp. 179–222.

Klemm, D. and Lehman, M. (2021) 'A prospective evaluation of treatment recommendations compared to outcomes for a lung cancer multidisciplinary team and legal implications', *Journal of Medical Imaging and Radiation Oncology*, 65(6), pp. 755–759. Available at: https://doi.org/10.1111/1754-9485.13192.

Klok, P.-J. and Denters, B. (2018) 'Structuring participatory governance through particular "rules in use": lessons from the empirical application of Elinor Ostrom's IAD Framework', *Handbook on Participatory Governance*, pp. 120–142.

Liu, X. *et al.* (2022) 'The medical algorithmic audit', *The Lancet Digital Health*, 4(5), pp. e384–e397. Available at: https://doi.org/10.1016/S2589-7500(22)00003-6.

Makoul, G. and Clayman, M.L. (2006) 'An integrative model of shared decision making in medical encounters', *Patient Education and Counseling*, 60(3), pp. 301–312. Available at: https://doi.org/10.1016/j.pec.2005.06.010.

Margetts, H., 2022. Rethinking AI for good governance. Daedalus, 151(2), pp.360-371.

McArdle, J. (2021) *Medical training pathway, The British Medical Association is the trade union and professional body for doctors in the UK*. Available at: https://www.bma.org.uk/advice-and-support/studying-medicine/becoming-a-doctor/medical-training-pathway

McMaster, E., Phillips, C. and Broughton, N. (2016) 'Righting the wrongs of traditional medical hierarchy', *Anaesthesia*, 71(1), pp. 110–111. Available at: https://doi.org/10.1111/anae.13352.

Methnani, L. *et al.* (2021) 'Let Me Take Over: Variable Autonomy for Meaningful Human Control', *Frontiers in Artificial Intelligence*, 4, p. 737072. Available at: https://doi.org/10.3389/frai.2021.737072.

Milanovich, D. *et al.* (1998) 'Status and cockpit dynamics: A review and empirical study', *Group Dynamics: Theory, Research, and Practice*, 2(3), pp. 155–167. Available at: https://psycnet.apa.org/doi/10.1037/1089-2699.2.3.155.

Morgan, D. *et al.* (2022) *High-stakes team based public sector decision making and AI oversight*. preprint. SocArXiv. Available at: https://doi.org/10.31235/osf.io/arq3w.

National Institute of Standards and Technology (2023). NIST AI 100-1. Artificial Intelligence Risk Management Framework (AI RMF 1.0). Available at: https://www.nist.gov/itl/ai-risk-management-framework/roadmap-nist-artificial-intelligence-risk-management-framework-ai.

National Institute of Standards and Technology (2023). AI RMF Playbook. Available at: https://airc.nist.gov/AI_RMF_Knowledge_Base/Playbook

NHS AI Lab & Health Education England (2022). *Developing healthcare workers' confidence in artificial intelligence (AI) (Part 2)*. Available at: https://digital-transformation.hee.nhs.uk/building-a-digital-workforce/dart-ed/horizon-scanning/developing-healthcare-workers-confidence-in-ai

NHS England and NHS Improvement (2019) *Shared Decision Making: Summary guide*. Available at: https://www.england.nhs.uk/wp-content/uploads/2019/01/shared-decision-making-summary-guide-v1.pdf

Oliver, D. (2017) 'David Oliver: Supervision and clinical autonomy for junior doctors—have we gone too far?', *BMJ*, p. j4659. Available at: https://doi.org/10.1136/bmj.j4659.

Ostrom, E. (1985) 'Formulating the elements of institutional analysis'. Available at: https://dlc.dlib.indiana.edu/dlc/handle/10535/2145.

Ostrom, E. (1986) 'An Agenda for the Study of Institutions', *Public Choice*, 48(1), pp. 3–25.

Ostrom, E. (2005) *Understanding institutional diversity*. Princeton, NJ: Princeton Univ. Press.

Ostrom, E. (2011) 'Background on the Institutional Analysis and Development Framework: Ostrom: Institutional Analysis and Development Framework', *Policy Studies Journal*, 39(1), pp. 7–27. Available at: https://doi.org/10.1111/j.1541-0072.2010.00394.x.

Paullada, A. *et al.* (2021) 'Data and its (dis)contents: A survey of dataset development and use in machine learning research', *Patterns*, 2(11), p. 100336. Available at: https://doi.org/10.1016/j.patter.2021.100336.

Peadon, R. (Rod), Hurley, J. and Hutchinson, M. (2020) 'Hierarchy and medical error: Speaking up when witnessing an error', *Safety Science*, 125, p. 104648. Available at: https://doi.org/10.1016/j.ssci.2020.104648.

Piciocchi, P. *et al.* (2019) 'Digital workers in service systems: challenges and opportunities', *Handbook of Service Science, Volume II*, pp. 409–432.

Radner, R. (1993) 'The Organization of Decentralized Information Processing', *Econometrica*, 61(5), p. 1109. Available at: https://doi.org/10.2307/2951495.

Raji, I.D. *et al.* (2022) 'The Fallacy of AI Functionality', in *2022 ACM Conference on Fairness, Accountability, and Transparency. FAccT '22: 2022 ACM Conference on Fairness, Accountability, and Transparency*, Seoul Republic of Korea: ACM, pp. 959–972. Available at: https://doi.org/10.1145/3531146.3533158.

Rashidi, H.H. *et al.* (2021) 'Machine learning in health care and laboratory medicine: General overview of supervised learning and Auto-ML', *International Journal of Laboratory Hematology*, 43(S1), pp. 15–22. Available at: https://doi.org/10.1111/ijlh.13537.

Rivera, S.C. *et al.* (2020) 'Guidelines for clinical trial protocols for interventions involving artificial intelligence: the SPIRIT-AI extension', *The Lancet Digital Health*, 2(10), pp. e549–e560. Available at: https://doi.org/10.1016/S2589-7500(20)30219-3.

Rudin, C. (2019) 'Stop explaining black box machine learning models for high stakes decisions and use interpretable models instead', *Nature Machine Intelligence*, 1(5), pp. 206–215. Available at: https://doi.org/10.1038/s42256-019-0048-x.

Sambasivan, N. *et al.* (2021) '"Everyone wants to do the model work, not the data work": Data Cascades in High-Stakes AI', in *Proceedings of the 2021 CHI Conference on Human Factors in Computing Systems. CHI '21: CHI Conference on Human Factors in Computing Systems*, Yokohama Japan: ACM, pp. 1–15. Available at: https://doi.org/10.1145/3411764.3445518.

Santoni de Sio, F. and van den Hoven, J. (2018) 'Meaningful Human Control over Autonomous Systems: A Philosophical Account', *Frontiers in Robotics and AI*, 5, p. 15. Available at: https://doi.org/10.3389/frobt.2018.00015.

Savioja, P. and Norros, L. (2013) 'Systems usability framework for evaluating tools in safety–critical work', *Cognition, Technology & Work*, 15(3), pp. 255–275. Available at: https://doi.org/10.1007/s10111-012-0224-9.

Schwartz, R. *et al.* (2022) 'Towards a Standard for Identifying and Managing Bias in Artificial Intelligence'. Available at: https://www.nist.gov/publications/towards-standard-identifying-and-managing-bias-artificial-intelligence (Accessed: 7 June 2022).

Simon, H.A. (1997) *Administrative behavior: a study of decision-making processes in administrative organizations*. 4th ed. New York: Free Press.

Social Impact Lab (SIMLab) (2017) SIMLab's framework for context analysis of inclusive technology in social change projects. Available at: http://simlab.org/resources/contextanalysis/

Smith, H. (2021) 'Clinical AI: opacity, accountability, responsibility and liability', *AI & SOCIETY*, 36(2), pp. 535–545. Available at: https://doi.org/10.1007/s00146-020-01019-6.

Smuha, N.A., (2021). Beyond the individual: governing AI's societal harm. Internet Policy Review, 10(3).

Sterz, S. *et al.* (2024) 'On the Quest for Effectiveness in Human Oversight: Interdisciplinary Perspectives'. arXiv. Available at: https://doi.org/10.48550/arXiv.2404.04059.

Straub, V.J., Morgan, D., Bright, J. and Margetts, H., 2023. Artificial intelligence in government: Concepts, standards, and a unified framework. *Government Information Quarterly*, *40*(4), p.101881.

Strauß, S. (2021) 'Deep Automation Bias: How to Tackle a Wicked Problem of AI?', *Big Data and Cognitive Computing*, 5(2), p. 18. Available at: https://doi.org/10.3390/bdcc5020018.

Swain, R. *et al.* (2021) '22 Evaluation and modification of medical handover using trainee feedback and technological advancements', in *Abstracts*. *Leaders in Healthcare 2021*, BMJ Publishing Group Ltd, p. A8.3-A9. Available at: https://doi.org/10.1136/leader-2021-FMLM.22.

Sydor, D.T. *et al.* (2013) 'Challenging authority during a life-threatening crisis: the effect of operating theatre hierarchy', *British Journal of Anaesthesia*, 110(3), pp. 463–471. Available at: https://doi.org/10.1093/bja/aes396.

Tenenberg, J. (2008) 'An institutional analysis of software teams', *International Journal of Human-Computer Studies*, 66(7), pp. 484–494. Available at: https://doi.org/10.1016/j.ijhcs.2007.08.002.

Tyworth, M., Giacobe, N.A., Mancuso, V.F., McNeese, M.D. and Hall, D.L., 2013. A human-in-the-loop approach to understanding situation awareness in cyber defence analysis. EAI Endorsed Transactions on Security and Safety, 1(2).

UK Foundation Programme (2021) 'Curriculum'. Available at: https://foundationprogramme.nhs.uk/curriculum/.
19